\documentclass[3p,times]{article}
\usepackage[a4paper,margin=3cm]{geometry}
\usepackage{stackengine,graphicx}
\usepackage{tikz,float}
\usetikzlibrary{shapes.geometric, arrows, positioning}
\usetikzlibrary{shapes,arrows,automata,positioning}
\usepackage{mathtools}
\usepackage{amsthm, amsmath, amsfonts, amssymb}
\usepackage[colorlinks=true, allcolors=blue]{hyperref}
\newcommand{\eps}{\epsilon}
\renewcommand{\frak}[1]{\mathfrak{#1}}
\newcommand{\code}[1]{\ulcorner #1 \urcorner} 
 
\newcommand{\pair}[1]{\langle #1 \rangle} 
\renewcommand{\tt}[1]{\texttt{#1}}
\newcommand{\bb}[1]{\mathbb{#1}}
\renewcommand{\cal}[1]{\mathcal{#1}}
\newcommand{\ov}[1]{\overline{#1}}

\theoremstyle{definition}
\newtheorem{thm}{Theorem}[section]
\newtheorem{defn}[thm]{Definition}

\newtheorem{prop}[thm]{Proposition} 
\newtheorem{rmrk}[thm]{Remark}
\newtheorem{qst}[thm]{Question}
\newtheorem{lemm}[thm]{Lemma}

\begin{document}
\title{On Positive Computation}
\author{Eduardo Skapinakis\\
{\small eduardo.skapinakis@uni-tuebingen.de}\\
{\small Carl Friedrich von Weizs{\"a}cker Zentrum, Universit{\"a}t T{\"u}bingen}\\
{\small Center for Mathematics and Applications (NOVA Math), NOVA FCT}
}
\maketitle
\begin{abstract}
We introduce new notions of reducibility for monotone classes and show that a monotone version of Circuit Satisfiability and the Hamiltonian Cycle Problem are complete for \tt{mNP}, even under \textit{positive} reductions (which are not the same as \textit{monotone} reductions). 
If this holds for all (monotone) \tt{NP}-complete problems -- that they are also so complete for \tt{mNP} --, then $\tt{P}\neq\tt{NP}$.
\,
We introduce positive models of relativized computation and show that the separation $\tt{posP} \subsetneq \tt{mNP}$ does not relativize, even when oracles are restricted to monotone sets. 
Yet, there is a (weak) diagonalization proof that $\tt{posP}\subsetneq\tt{mNP}$. \\[0.1cm]
\end{abstract}
\section{Introduction}
Monotone Complexity is a branch of Complexity Theory focusing on computational models to which \textit{monotonicity} restrictions are imposed. Starting with Razborov's lower bounds on monotone circuits \cite{razborov:85,razborov:85b}, a variety of results in monotone Complexity have been proven \cite{tardos:88,karchmer:88,raz:97}, whose non-monotone counterparts remain open problems (see e.g. \cite[Section $2.6$]{grigni:92} for an overview).

While the usual monotone computational model is a $\neg$-free circuit, there are also \textit{uniform} models, as \textit{positive Turing machines} \cite{grigni:92,lautemann:96,lautemann:98,das:18,skapinakis:26}. These are the focus of this paper.
Based on the machine model defined in \cite{das:18}, we introduce two \textit{new} notions of reducibility for monotone classes: \textit{positive many-to-one} reducibility and \textit{positive Turing} reducibility.

We give a simple quantifier characterization of \tt{mNP} and prove that a monotone version of Circuit Satisfiability and the Hamiltonian Cycle Problem are complete for \tt{mNP}, even when reductions are restricted to being \textit{positive} (which is not the same as \textit{monotone} \cite{razborov:85,razborov:85b}).
In both cases, the reductions constructed are simple and leverage the fact that parts of known reductions depend only on the \textit{size} of the input.
Interestingly, if this holds for all \tt{NP} complete problems that are monotone, i.e., that they are also \tt{mNP} complete under \tt{posFP} reductions, then $\tt{P}\neq\tt{NP}$.

We then construct a \textit{monotone} oracle $A$ for which $\tt{posP}^A = \tt{mNP}^A$, showing that the separation of these classes does not relativize, even when oracles are restricted to being monotone.
Adapting the work of Kozen \cite{kozen:78}, we show also that there exists a weak diagonalization proof that $\tt{posP}\neq\tt{mNP}$, which we discuss in the context of diagonalization versus relativizing proof techniques.
\section{The complexity classes}
All necessary background from Complexity can be found in \cite{balcazar:88,arora:16}. 
Let $\bb{W} = \{0,1\}^*$ be the set of $0$-$1$ words.
Given $x\in\bb{W}$, $|x|$ denotes the number of bits of $x$ and, for $0\leq i<|x|$, $x(i)$ denotes the $i$th bit of $x$, counting from the right (so $x = x(|x|-1)\cdots x(1)x(0)$).\footnote{This notation avoids confusing different words $x_0,x_1\in\bb{W}$ with the bits $0$ and $1$ of $x$, denoted by $x(0)$ and $x(1)$.} If $i<0$ or $i\geq |x|$, then $x(i) := \eps$.
We abbreviate $x_1,\dots,x_k\in\bb{W}$ as $\ov{x}\in\bb{W}^k$, and denote $(|x_1|,\dots,|x_k|)$ by $|\ov{x}|$.

We consider Turing machines over the alphabet $\Sigma = \{\triangleright,\square,1,0\}$.
We call $\Box$ the ``blank'', and $\triangleright$ the ``beginning of the tape marker''

\begin{defn}\label{def:turing_machine}
A $k$-tape DTM is a tuple $M=(Q,\delta,s,h)$ where:
\begin{itemize}
    \item[] $Q$ is a finite set of non-final states. $s\in Q$ is the starting state and $h\notin Q$ the halting state
    \item[] $\delta: Q\times \Sigma^k \rightarrow (Q\cup\{h\})\times (\Sigma\times\{L ,S ,R\} )^k$ is the transition function\\
    $L ,S $ and $R$ represent the head-moves ``left'', ``stay'' and ``right''.
\end{itemize}
We assume that $M$ always moves right when it reads the symbol $\triangleright$: 
whenever $\delta (q,\sigma_1,\cdots,\sigma_k)=(q',\tau_1, d_1,\cdots,\tau_k,d_k)$, if $\sigma_j = \triangleright$, then $\tau_j =\triangleright$ and $d_j =R$. 
\end{defn}

Note that there is no accepting nor rejecting state in the definition above. A DTM rather has a designated output tape, and outputs $0$ or $1$ to indicate ``rejection'' or  ``acceptance''. This choice is necessary for Theorem \ref{thm:posDTM_circuits} to hold (see Appendix \ref{sec:no_accepting_states}). 
We may denote by $M(x_1,\cdots,x_n)$ the output of a machine $M$ when given $x_1,\cdots,x_n$ as input.

\begin{defn}\label{def:basic_classes}
\tt{FP} is the set of functions computable in polynomial time by a DTM. \tt{FPSPACE} is the set of functions computable in polynomial space by a DTM.
\end{defn}

Since any binary function can be seen as the characteristic function of some set, we will often identify classes of binary functions with classes of sets.

\begin{defn}\label{def:basic_binary_classes}
\tt{P} is the set of binary \tt{FP} functions, or sets whose characteristic function is in \tt{FP}. 
\tt{PSPACE} is the set of binary \tt{FPSPACE} functions, or sets whose characteristic function is in \tt{FPSPACE}.
\end{defn}

A non-deterministic Turing machine (NTM) is defined similarly to its deterministic version, except that its transition is a \textit{relation}, which, given a state and the scanned bits, may return a set of possible next configurations.
For simplicity, we may assume that $\delta$ returns precisely \textit{two} next configurations, whence we may represent the transition relation as a pair of two transition functions $\delta = (\delta_0,\delta_1)$. Given an input $x$, these functions generate a computation tree.
A NTM outputs $0$ and $1$ instead of rejecting and accepting, with the convention that it outputs $1$ iff there is some branch in its computation tree outputting $1$.

\begin{defn}\label{def:NP}
\tt{NP} is the set of binary functions computable in polynomial time by a NTM.    
\end{defn}
\subsection{Monotone complexity}\label{def:monotone_complexity}
Define the partial order $\leq$ on $\bb{W}$ as follows: 
$x\leq y$ iff $|x| = |y|$ and, for $k=0,\cdots,|x|-1$, if $x(k) = 1$, then $y(k) = 1$.
For example, $001\leq 101$, but $001\not\leq 100$.
A function on $\bb{W}$ is \textit{monotone} if, whenever $x\leq y$, $f(x)\leq f(y)$.
If $f$ is the \textit{monotone} characteristic function of a set $A\subseteq\bb{W}$, then, if $x\in A$ and $x\leq y$, it follows that $y\in A$. 

\begin{defn}\label{def:circuits}
A monotone (or $\neg$-free) circuit is a circuit consisting only of input gates, constant gates ($0$ and $1$), $\wedge$ and $\vee$ gates, and an output gate, which, by convention, is the circuit's last gate.
A family of circuits $\{C_n\}_{n\geq 1}$ is \tt{P}-uniform if the function $1^n\mapsto C_n$, which, given a unary word of size $n$, returns a code of $C_n$, is computable in polynomial time.
A binary function $F$ is computed by a circuit family $\{C_n\}$ if, for all $x\in\bb{W}$, $F(x)=C_{|x|}(x)$.
\end{defn}

Being monotone is a semantic property, it does not depend on the machinery computing $f$.
On the other hand, $\neg$-free circuits, which only compute monotone functions, have an intrinsic monotonicity restriction.
To distinguish these two notions we follow \cite{lautemann:96,lautemann:98,das:18} and reserve the word \textit{monotone} for the semantic level, and say that a function is \textit{positive}, if it can be computed by a restricted model of computation.

\begin{defn}\label{def:mC}
Let $\cal{C}$ be a complexity class of function (or sets, identified by its characteristic function).
$\tt{m}\cal{C}$ denotes the class of functions (or sets) in $\cal{C}$ that are monotone.
$\tt{pos}\cal{C}$ denotes the positive counterpart of $\cal{C}$, i.e., the class obtained by imposing the same resource restriction, with the additional condition that the computational model is restricted to be positive (e.g. a $\neg$-free circuits).
\end{defn}

If \tt{pos}$\cal{C}$ and $\tt{m}\cal{C}$ coincide, we use the latter notation and say that $\tt{m}\cal{C}$ is \textit{trivial}.

\begin{defn}[\cite{grigni:92}]\label{def:classes}
\tt{posP} is the class of languages decidable by \tt{P}-uniform monotone circuits;
\tt{mNP} is the class of languages decidable by \tt{P}-uniform non-deterministic monotone circuits.\footnote{Monotone non-deterministic circuits are allowed to have negation gates, as long as they only depend on the non-deterministic input variables.}
\end{defn}
 
The class $\tt{mNP}$ is trivial \cite{grigni:92}, but $\tt{posP}$ is not \cite{razborov:85,razborov:85b}. 
In particular, $\tt{posP}\subsetneq \tt{mP}\subseteq \tt{mNP}$. 
This is one of the motivations for interpreting monotone circuit lower bounds as separations of complexity classes \cite{grigni:92}: they are \textit{monotone analogues} of separations believed to be true.
\section{Positive Turing machines}\label{def:positive_functions}
To define positive deterministic Turing machines (posDTMs), define the following ordering on configurations and instructions:

\begin{defn}[\cite{das:18}]\label{def:order}
Denote $Q\times \Sigma^k$ by $\frak{I}$ and $(Q\cup\{h\})\times (\Sigma\times\{L ,S ,R\} )^k$ by  $\frak{O}$. Define the partial orders $\preceq_\frak{I}$ and $\preceq_\frak{O}$ as follows: 
\begin{itemize}
    \item[] $(q,\sigma_1,\cdots,\sigma_k)\preceq_\frak{I} (q',\sigma'_1,\cdots,\sigma'_k)$ if: $q=q'$ and, for $1\leq j\leq k$, either $\sigma_j =\sigma'_j$, or both $\sigma_j =0$ and $\sigma'_j=1$
    \item[] $(q,\tau_1, d_1,\cdots,\tau_k,d_k)\preceq_\frak{O} (q',\tau'_1, d'_1,\cdots,\tau'_k,d'_k)$ if: $q=q'$ and, for $1\leq j\leq k$, $d_j=d'_j$ and either $\tau_j =\tau'_j$, or both $\tau_j =0$ and $\tau'_j=1$
\end{itemize} 
\end{defn}

\begin{defn}[\cite{das:18}]\label{positive_DTM}
A $k$-tape posDTM is a $k$-tape DTM $M=(Q,\delta,s,h)$ for which $\delta: \frak{I}\rightarrow\frak{O}$ is monotone with respect to the partial orders $\preceq_\frak{I}$ and $\preceq_\frak{O}$. 
That is, whenever $I\preceq_\mathfrak{I} I'$, it holds that $ \delta(I)\preceq_\mathfrak{O} \delta(I')$.
\end{defn}

Whenever a posDTM, in state $q$, writes a $1$ when scanning the bit $0$, it must also write a $1$ when scanning the bit $1$. Moreover, the next state and direction followed by the machine cannot depend on whether the scanned bit is $0$ or $1$: posDTM are \textit{oblivious}.
Regarding the sets decidable by a posDTM, we have the following equivalence:

\begin{thm}[\cite{das:18}]\label{thm:posDTM_circuits}
The following are equivalent:
\begin{itemize}
    \item[(i)] binary functions computable by posDTMs that halt in polynomial time
    \item[(ii)] binary functions computable by $\tt{P}$-uniform circuit families of $\neg$-free circuits
\end{itemize}
\end{thm}

The important point of this theorem is the uniformity condition. Since the \tt{FP}-function generating the code of each circuit $C_n$, only receives $1^n$ as input, it only has to use the \textit{size} of the input received, and thus can be implemented by a \textit{positive} Turing machine.

\begin{defn}[\cite{das:18}]\label{def:posFP}
\tt{posFP} is the class of functions that can be computed by posDTMs in polynomial time.
\tt{posP} is the class of binary \tt{posFP} functions.
\end{defn}
\subsection{Positive pairing and unpairing}\label{sec:positive_pairing}
We will require positive pairing and unpairing functions. Note that these are usually defined by a case distinction on a \textit{separation marker} (usually $01$), which is not a positive operation. 
We start with some auxiliary functions.

The following functions are in \tt{posFP}:
\begin{itemize}
    \item[] $\oplus:\bb{W}^2\to\bb{W}$, such that $x\oplus y = xy$ (word concatenation)
    \item[] $\vee:\bb{W}^2\to\bb{W}$, such that $x \vee y = z$, where $|z|=\max\{|x|,|y|\}$ and $z(k) = \max\{x(k), y(k)\}$, for $k < |z|$ (join)\footnote{By convention, $\max\{\eps,i\} = \max\{i,\eps\} = i$, for $i\in\{0,1\}$. For example, $10\vee000 = 010$.}
    \item[] $\tt{CANTOR}:\bb{W}^2\to\{0\}^*$, such that $|\tt{CANTOR}(x,y)|=(|x|+|y|)(|x|+|y|+1)/2+|x|$
    \item[] $\tt{P1}:\bb{W}\to\{1\}^*$, such that $|\tt{P1}(\tt{CANTOR}(x,y))|=|x|$
    \item[] $\tt{P2}:\bb{W}\to\{1\}^*$, such that $|\tt{P2}(\tt{CANTOR}(x,y))|=|y|$
\end{itemize}

Now, the positive pairing function is defined so that the \textit{size} of the word replaces the role of a separation marker, allowing a positive function to calculate the size of the paired words simply by the size of their pairing.

\begin{defn}\label{def:pairing}
Define the following positive pairing and unpairing functions:
\begin{itemize}
    \item[] $\pair{\,\cdot\,,\,\cdot\,}:\bb{W}^2\to\bb{W}$, such that $\pair{x,y} = (x\oplus y)\vee \tt{CANTOR}(x,y)$ (pairing)
    \item[] $(\,\cdot\,)_1:\bb{W}\to\bb{W}$, such that $(z)_1$ returns the last $\tt{P2}(z)$ bits of $z$ (right unpairing)
    \item[] $(\,\cdot\,)_0:\bb{W}\to\bb{W}$, such that $(z)_0$ returns the last $\tt{P1}(z)$ bits of $z$, after the last $\tt{P2}(z)$ bits have been removed (left unpairing)
\end{itemize}
\end{defn}

For example, $\pair{1,01} = (1\oplus 01)\vee 0^{7} = 0000101$. From the size of $\pair{1,01}$, \textit{seven}, one can calculate the size of the inputs being paired: \textit{one} and \textit{two}, respectively.
Thus, even though the function $(x,y)\mapsto\pair{x,y}$ is not surjective, and the functions $z\mapsto(z)_i$ not injective, together they satisfy the property: 
for $i\in\{0,1\}$ and for $x_0,x_1\in\bb{W}$, $(\pair{x_0,x_1})_i = x_i$.
By convention, $\pair{x_0,\dots,x_n}$ abbreviates $\pair{x_0,\pair{x_1,\dots}\dots}$.
\subsection{Monotone circuits}\label{sec:circuits}
We start by fixing an encoding for monotone circuits such that \textit{circuit satisfiability} is a monotone problem.
Recall that a circuit with $n$ input gates has two further constant gates, representing $0$ and $1$.

\begin{defn}\label{def:encoding_monotone_circuit}
A $\neg$-free circuit with $n$ inputs and $k$ gates is encoded as a word
\begin{align*}
    &\underbrace{w_0^1w_0^{1'}}_{\mathclap{\text{Type of gate}}}w_1^1\dots w_{n+2}^1\oplus w_0^2w_0^{2'}\underbrace{w_1^2\dots w_{n+2}^2}_{{\mathclap{\text{Connections to the input and constant gates}}}}w_{n+3}^2\oplus\cdots\oplus w_0^kw_0^{k'}w_1^k\dots w_{n+2}^k\underbrace{w_{n+3}^k\dots w_{n+1+k}^k}_{\mathclap{\text{Connections to the previous gates}}}&    
\end{align*}
of size $2k + (n+2)k + k(k-1)/2$, where:
\begin{itemize}
    \item[] for $1\leq m\leq k$, $w_0^mw_0^{m'}$ is $01$ if $m$ is an $\wedge$ gate and $10$ if it is an $\vee$ gate
    \item[] for $1\leq m\leq k$, the bits $w_1^m\dots w_{n+2+(m-1)}^m$ establish the connections of the $m$th gate to the previous gates (including the input and constant gates) as follows:
    \begin{itemize}
        \item[] if $m$ is a $\wedge$ gate, $w_i^m=0$ indicates a connection and $w_i^m=1$ indicates \textbf{no} connection
        \item[] if $m$ is a $\vee$ gate, $w_i^m=0$ indicates \textbf{no} connection and $w_i^m=1$ indicates a connection
    \end{itemize}
\end{itemize}
\end{defn}

As an example, the word $01\oplus 110\oplus 10$ represents a $\wedge$ gate connected to $x(0)$ and to the constant $1$. 
Given a $\neg$-free circuit $C$, denote by $\code{C}$ its code under the above definition.

\begin{prop}\label{prop:eval_circuit}
There is a \tt{posFP} function $\tt{EVAL}:\bb{W}^2\to\{0,1\}$ such that, if $C$ is a circuit with $n$ input gates and $|x|=n$, $\tt{EVAL}(\code{C},x) = C(x)$.
\end{prop}
\begin{proof}
We describe a \tt{posFP} machine $M$.
Assume $M$ has a work tape where it will store the value of each calculated gate.
We show how to compute the value of the first gate. The remaining are computed analogously using the bits stored in the work tape.

Let $e,x\in\bb{W}$.
From the size of $|x|:=n$, $M$ starts by reading the first $n+4$ bits of $e$, obtaining a block $w_0^1w_0^{1'}w_1^1\dots w_{n+2}^1$ (this will only be correct when indeed $e=\code{C}$ has $|x|$ input gates).
Let
\begin{align*}
    &\tt{EVALAND}_1(e,x) = \underbrace{(x(0)\vee w_n^1)}_{\mathclap{\text{$w_n^1 = 0$ indicates a connection with $x(0)$, the rightmost input bit}}}\wedge (x(1)\vee w_{n-1}^1)\wedge \dots\wedge(x(n-1)\vee w_1^1)\wedge \underbrace{w_{n+1}^1}_{\mathclap{\text{Connection to the constant } 0}}&\\
    &\tt{EVALOR}_1(e,x) = \underbrace{(x(0)\wedge w_n^1)}_{\mathclap{\text{$w_n^1 = 1$ indicates a connection with $x(0)$}}}\vee (x(1)\wedge w_{n-1}^1)\vee \dots\vee(x(n-1)\wedge w_1^1)\vee \underbrace{w_{n+2}^1}_{\mathclap{\text{Connection to the constant } 1}}&
\end{align*}
Then, $M$ computes: 
\begin{align*}
    &\tt{EVAL}_1(e,x) = (\underbrace{w_0^{1'}}_{\mathclap{\text{Is $1$ for $\wedge$ gates, and $0$ for $\vee$ gates}}}\wedge \tt{EVALAND}_1(e,x))\vee (\tt{EVALOR}_1(e,x) \wedge \underbrace{w_0^1}_{\mathclap{\text{Is $0$ for $\wedge$ gates, and $1$ for $\vee$ gates}}})&
\end{align*}
If $e$ really encodes a circuit, then $w_0^1w_0^{1'}\in\{01,10\}$.
If $w_0^1w_0^{1'} = 01$, then only the first disjunct remains, which computes a $\wedge$ gate.
Indeed 
\begin{align*}
    \tt{EVALAND}_1(e,x)&= \bigwedge_{i=0}^{n-1} (x(i)\vee w_{n-i}^1) \wedge w_{n+1}^1 = \bigwedge_{0\leq i\leq n-1\,:\,w_{n-i}^1=0} x(i) \wedge w_{n+1}^1 &
\end{align*}
Since a connection from a $\wedge$ gate is identified by a $0$, the above expression represents the computation of a $\wedge$ gate. Note that if there is a connection to the constant $0$, $w_{n+1}^1 = 0$ and so the whole expression evaluates to $0$.
Analogously, if $w_0^1w_0^{1'} = 10$, only the second disjunct remains, which computes a $\vee$ gate.

This concludes the evaluation of the first block.
Now the later blocks are evaluated with the extra task of searching the work tape for the values of the previous gates. In the end, the machine outputs the value of the last gate.
As there are $k$ gates, this operation will have to be performed at most $k$ times, which yields an algorithm that is polynomial in $|e|$ and $|x|$.
\end{proof}

Note that the function \tt{EVAL} does not test if $e$ correctly encodes a circuit, nor does it try to calculate the number of input gates of a circuit from its code.
This would require a case distinction on bits, which is a non-monotone operation.
Thus, \tt{EVAL} will only behave as desired whenever $e=\code{C}$, for $C$ a circuit with $|x|$ input bits.

Now the monotone circuit value problem (\tt{mCVP}) consists of the pairs $\code{C}$ (circuit) and $x$ (input), such that $\tt{EVAL}(\code{C},x) = 1$.

\begin{defn}\label{def:mCVP}
$\tt{mCVP} := \{x\in\bb{W}\,|\, \tt{EVAL}((x)_0,(x)_1)=1\}$.
\end{defn}

From Proposition \ref{prop:eval_circuit}, $\tt{mCVP}\in \tt{posP}$.
Moreover, \tt{mCVP} is complete for \tt{P} under (non-monotone) \tt{logspace} reductions \cite{goldschlager:77}. 
\subsection{Pushing negations to the input level}\label{sec:pushing_negations}
Similarly to how DeMorgan's laws can be applied \textit{efficiently} to a circuit, to ``push negations'' to the input level, a similar phenomenon can be proven for positive functions.

Define the \tt{FP} function $\diamond :\bb{W}\to\bb{W}$ as: $\diamond(\eps) = \eps$, $\diamond(xi) = \diamond(x)\oplus i\oplus\neg i$, for $i\in\{0,1\}$.
Given $x\in\bb{W}$, $\diamond(x)$ is the \textit{interleaving} of $x$ with its negation (e.g. $\diamond(10) = 1001$).
When no confusion arises, we will simply denote $\diamond(x)$ by $\diamond x$.

\begin{defn}\label{def:predictable}
Call a function $F:\bb{W}^k\to\bb{W}$ \textit{predictable} if there is $F'\in\tt{posFP}$ such that, for every $\ov{x}\in\bb{W}^k$, $|F(\ov{x})|=|F'(\ov{x})|$.
\end{defn}

The following is stated for predictable functions, but will in practice only be used for functions whose size is precisely a monotone polynomial.

\begin{thm}\label{thm:diamond}
Let $F:\bb{W}^k\to\bb{W}$ be an \tt{FP} \textit{predictable} function. 
Then, there exists a \tt{posFP} function $f:\bb{W}^k\to\bb{W}$ such that, for all $\ov{x}\in\bb{W}^k$, $F(x_1,\dots,x_k)=f(\diamond x_1,\dots,\diamond x_k)$.
\end{thm}
\begin{proof}
We prove that the result holds for binary functions $F:\bb{W}\to\{0,1\}$. Then, the output of a \textit{predictable} function can be obtained by calculating its bits and concatenating them.

Let $F:\bb{W}\to\{0,1\}$ in \tt{FP}.
Since the monotone circuit value problem is complete for \tt{P} under logspace reductions \cite{goldschlager:77}, there is a \tt{P}-uniform (indeed \tt{L}-uniform) monotone circuit family $\{C_n\}$ such that: for each $x\in\bb{W}^n$, $C_{2n}(x(n-1)\neg x(n-1)\cdots x(0)\neg x(0)) = F(x)$ (this requires only a re-ordering of the input gates from the construction in \cite{goldschlager:77}).

The function $x\mapsto C_{|x|}(x)$ is computed by a \tt{P}-uniform monotone circuit family, and so there is a \tt{posFP} function $f$ computing it (Theorem \ref{thm:posDTM_circuits}).
Thus, $F(x)=f(\diamond x)$ as desired.
\end{proof}

This result gives us some freedom to construct \tt{posFP} reductions amongst different problems. Usually, part of the instances constructed in a reduction do not actually depend on the input, but only on its size (see Section \ref{sec:positive_many-to-one_completeness} for two examples). For these parts, one can construct their negation as well and use the theorem above.

Note that, even when $\diamond$ is not applied to all inputs, one can still construct positive functions which behave negatively on those inputs, for which their negation is also given.

\begin{lemm}[\cite{skapinakis:26}]\label{lemm:leq_diamond} 
There is a \tt{posFP} function $\chi_{\leq}$ such that, for every $x,y\in\bb{W}$ such that $|x|=|y|$: $\chi_{\leq}(\diamond x,y) = 1$ iff $x\leq y$    
\end{lemm} 
\begin{proof}
Consider the \tt{posP} function $(z,y)\mapsto \bigwedge_{i=0}^{|y|-1}(z(2i)\vee y(i))$.
Whenever $z = \diamond x$, the bits $z(2i)$ correspond to $\neg x(i)$.
Thus, for each $x,y\in\bb{W}$, $\chi_{\leq}(\diamond x,y) = \bigwedge_{i=0}^{|y|-1}(\neg x(i)\vee y(i))$, which is $1$ iff, for each bit position $i$ where $\neg x(i) = 0$, it holds that $y(i) = 1$. 
Thus, for each $x,y\in\bb{W}$ such that $|x|=|y|$, $\chi_{\leq}(\diamond x,y) = 1$ iff, for each $0\leq i<|x|$, $x(i)\leq y(i)$.
\end{proof}

Note that $x\leq y$ is not monotone on $x$, but on $y$, which is why only $\diamond x$ is required.
\section{Monotone \tt{NP}}
A natural candidate for a \tt{mNP} problem would be a restriction of circuit satisfiability (\tt{CSAT}) \cite{cook:71,levin:73}, where circuits are required to be monotone.
However, if a circuit $C$ has no negations, then it is satisfiable iff $C(1^n) = 1$, so this variation of \tt{CSAT} is in \tt{posP}.
This has already been remarked in \cite{grigni:92}: \tt{mNP} requires the presence of \textit{controlled} negative operations. 
Based on the fact that $\tt{mNP} = \tt{NP}\cap\tt{mono}$ \cite[Theorem $2.2$]{grigni:92} and Theorem \ref{thm:diamond}, we give a quantifier description of \tt{mNP}.

\begin{thm}\label{thm:logical_characterization}
A set $A$ is in $\tt{mNP}$ iff there is a monotone polynomial $q$ and a function $F\in\tt{posP}$ such that:
$x\in A$ iff $\exists z\in\{0,1\}^{q(|x|)}\,:\,F(x,\diamond z) = 1$.    
\end{thm}
\begin{proof}
Since sets satisfying the stated condition are monotone, and in \tt{NP}, they are in \tt{mNP}.
Now, consider $A\in\tt{mNP}$. 
Since, in particular, $A\in\tt{NP}$, there is a function $F\in\tt{P}$ and a monotone polynomial $q$ such that $x\in A$ iff $\exists w\in\{0,1\}^{q(|x|)}\,:\,F(x,w)=1$.
Since $A$ is monotone, $x\in A$ iff there is $y\leq x$ such that $y\in A$.
From Theorem \ref{thm:diamond}, let $F'$ be the \tt{posP} function such that, for all $y,w\in\bb{W}$, $F'(\diamond y,\diamond w) = F(y,w)$.
Then, $x\in A$ iff there is $y\leq x$ and $w\in\{0,1\}^{q(|x|)}$ such that $F'(\diamond y,\diamond w) = 1$.

Let
\begin{align*}
    &F''(x,z) := \chi_\leq\bigg(\underbrace{z(|z|-1)\cdots z(|z|-2|x|)}_{\text{Represents }\diamond y}\,,\,x\bigg)\wedge&\\
    &\qquad\qquad F'\bigg(z(|z|-1)\cdots z(|z|-2|x|)\,,\,\underbrace{z(|z|-2|x|-1)\cdots z(0)}_{\text{Represents }\diamond w}\bigg)&
\end{align*}
Given $y\in\{0,1\}^{|x|}$ and $w\in\{0,1\}^{q(|x|)}$, $F''(x,\diamond y\oplus \diamond w) = \chi_\leq(\diamond y,x)\wedge F'(\diamond y,\diamond w)$.
Then, $x\in A$ iff there is $z\in\{0,1\}^{|x|+q(|x|)}$ such that $F''(x,\diamond z)=1$.    
\end{proof}

The advantage of this description is that one does not have to make a distinction on types of inputs to $F$, to allow only negative operations on the non-deterministic bits.
The negative operations are \textit{pushed} to the existential quantifier.\footnote{Alternatively, the acceptance condition could be stated without $\diamond$, but as follows: $x\in A$ iff there is $z\in\{0,1\}^{q(|x|)}$ such that $F(x,z) = 1$, $|z|$ is even, and, for $0\leq i \leq |z|/2-1$, $z(2i+1) = \neg z(2i)$. This way, the only negative operations are placed on \textit{checking} that $z = \diamond w$, for some $w\in\bb{W}$.}
Note also that the only place where the actual value of $x$ is used, is in computing $\chi_\leq$, to check that $y\leq x$. The remaining part depends only on the \textit{size} of $x$.
This characterization motivates the following definition:

\begin{defn}[\cite{skapinakis:26}]\label{def:positive_non_det}
A $k$-tape posNTM is a tuple $M = (Q,\delta, s,h)$, with $Q,s$ and $h$ defined as in Definition \ref{def:turing_machine}, and such that $\delta = (\delta_0,\delta_1)$ is a transition relation with both $\delta_0$ and $\delta_1$ \textit{monotone} functions, with respect to the orders defined in Definition \ref{def:order}. 
\end{defn}

Having two \textit{monotone} transition functions, a posNTM can have the function $\delta_0$ writing a $0$ and then a $1$, on the work tape, and the function $\delta_1$ writing a $1$ and then a $0$. 
This corresponds to having $\delta_0$ ``guessing'' $01$ and $\delta_1$ ``guessing'' $10$, which allows a posNTM to represent a quantifier of the form given in Theorem \ref{thm:logical_characterization}.
Thus, one obtains:

\begin{thm}[\cite{skapinakis:26}]\label{thm:mNP_new_machine_model}
A set is in $\tt{mNP}$ iff it is decidable by a posNTM in polynomial time.
\end{thm}
\section{Monotone \tt{NP}-completeness}\label{sec:positive_many-to-one_completeness}
Monotone completeness has been studied before, through a very weak form of reduction, called monotone projection translation \cite{grigni:92,stewart:94,stewart:95} (see also \cite{stewart:93}).
We study \tt{mNP} completeness using \tt{posFP} reductions.

\begin{defn}\label{def:completeness_posP}
Let $\cal{C}$ be a complexity class and $A\in\cal{C}$.
$A$ is \textit{complete} for $\cal{C}$ under \tt{posFP} many-to-one reductions if:
for every $B\in\cal{C}$ ,there is $F\in\tt{posFP}$ such that $x\in B$ iff $F(x)\in A$.
We abbreviate the latter condition to $B\leq_m^{pos} A$.
\end{defn}

Consider the following questions: 
\begin{qst}\label{qst:mNP_complete}
If $A\in\tt{mNP}$ and $A$ is \tt{NP}-complete under many-to-one \tt{FP} reductions, is $A$ also \tt{mNP}-complete under \tt{posFP} reductions?
\end{qst}

It seems to be a simple question. Yet, if true, it implies that $\tt{P}\neq\tt{NP}$. Before proving this (Theorem \ref{thm:p_neq_np}), we show that it still holds when we allow certain \textit{re-encodings}.

\begin{defn}\label{def:positive_encoding}
Call a function $\Xi:\cal{P}(\bb{W})\to\cal{P}(\bb{W})$ a \textit{positive} encoding if $\{\Xi(A):A\in\tt{posP}\}\subseteq\tt{posP}$.
\end{defn}

\begin{prop}\label{prop:positive_encoding}
There is a positive encoding $\Xi:\cal{P}(\bb{W})\to\cal{P}(\bb{W})$ such that: if $A$ is \tt{NP}-complete, $\Xi(A)$ is \textit{monotone} and complete for \tt{NP} under \textit{predictable} reductions.
\end{prop}
\begin{proof}
Given $A$ \tt{NP}-complete, define $A'=\{x10^*\,:\,x\in A\}$ and $\Xi(A)$ such that
\begin{align*}
    &1_{\Xi(A)}(x) = 
    \begin{cases}
        0 & \text{ if $|x|$ is odd or there is $i\leq |x|/2-1$ such that $x(2i) = x(2i+1) = 0$}\\
        1 & \text{ if there is $i\leq |x|/2-1$ such that $x(2i) = x(2i+1) = 1$}\\
        1_{A'}(\tt{ODD}(x)) & \text{ otherwise}
    \end{cases}&
\end{align*}
where $\tt{ODD}(x)$ returns the bits of $x$ that are in odd numbered positions.
Due to the first two conditions, whenever $x\leq y$ (for $|x|$ even) and $x\neq y$, either $x\notin \Xi(A)$ (if there is $i$ such that $x(2i) = x(2i+1) = 0$) or $y\in \Xi(A)$ (if there is $i$ such that $y(2i) = y(2i+1) = 1$). Thus, $\Xi(A)$ is monotone.

Now, let $C\in\tt{NP}$ and let $F\in\tt{FP}$ be a reduction from $C$ to $A$, such that $|F(x)|\leq p(|x|)$, for some monotone polynomial $p$.
Let $F'(x):=F(x)\oplus 10^{m(x)}$, where $m(x) = p(|x|)+1-|F(x)|$.
Then, $x\in C$ iff $F'(x)\in A'$ iff $\diamond F'(x)\in\Xi(A)$ and $|F'(x)| =  |F(x)| + 1 + p(|x|)+1-|F(x)| =  p(|x|)+2$, for every $x\in\bb{W}$, so the function $x\mapsto \diamond F'(x)$ is a predictable reduction from $C$ to $\Xi(A)$.

\smallskip

Now, we must prove that $\Xi$ is a positive encoding. 
Let $A\in\tt{posP}$. Then, $A'$, as defined above, is in $\tt{P}$.
Thus, there is a \tt{posP} function $F$ such that, for all $w\in\bb{W}$, $1_{A'}(w) = F(\diamond w)$.
Now, $\Xi(A)$ is decided by the \tt{posP} function $G$ which, given $x$, returns $0$ if $|x|$ is odd, and otherwise computes:
\begin{align*}
	&\bigg(\bigvee_{i=0}^{|x|/2-1}(x(2i)\wedge x(2i+1))\,\,\vee\,\,F(x) \bigg)\,\wedge\,\bigwedge_{i=0}^{|x|/2-1}(x(2i)\vee x(2i+1)&
\end{align*}
Let $x\in\bb{W}$.
If $|x|$ is odd, $G(x)=0$.
If there is $i\leq |x|/2-1$ such that $x(2i) = x(2i+1) = 0$, the right conjunction will return $0$, so $G(x)=0$. Otherwise, it will return $1$.
If there $i\leq |x|/2-1$ such that $x(2i) = x(2i+1) = 1$, then the big disjunction returns $1$. Since the right conjunction returns $1$ as well, $G(x) = 1$. 
Finally, the only case missing is when $x=\diamond w$, for $w\in\bb{W}$. In this case, $w = \tt{ODD}(x)$.
From the equality $1_{A'}(w) = F(\diamond w)$, it follows that $G(x) = F(\diamond w) = 1_{A'}(w) = 1_{A'}(\tt{ODD}(x)) = 1_{\Xi(A)}(x)$.

Thus, $\Xi(A)$ is decidable by $G$, and is thus in $\tt{posP}$, so $\Xi$ is a positive encoding.
\end{proof}

This result allows us to relax the conditions of Question \ref{qst:mNP_complete}, and still obtain the same conclusion.

\begin{thm}\label{thm:p_neq_np}
Suppose that for every monotone problem $A$ that is \tt{NP}-complete, under \tt{FP} reductions, there is a positive encoding $\Xi$ such that $\Xi(A)$ is \tt{mNP}-complete, under \tt{posFP} reductions.
Then $\tt{P}\neq\tt{NP}$.
\end{thm}
\begin{proof}
We prove the contrapositive.
Suppose that $\tt{P} = \tt{NP}$. 
Then, \tt{mCVP} is \textit{monotone} and \tt{NP}-complete under \tt{FP} reductions.
Now, let $\Xi$ be any positive encoding.
Then, $\Xi(\tt{mCVP})\in\tt{posP}$, so, if $A\leq_m^{pos}\Xi(\tt{mCVP})$, $A\in\tt{posP}$.
As $\tt{posP}\subsetneq\tt{mNP}$, this implies that $\Xi(\tt{mCVP})$ cannot be \tt{mNP}-complete under \tt{posFP} reductions.
Thus, there is no \textit{positive encoding} for which \tt{mCVP} is \tt{mNP}-complete under \tt{posFP} reductions.
\end{proof}

\begin{rmrk}\label{rmrk:almost_positive_reduction}
Suppose that $A\in\tt{mNP}$ is \tt{NP}-complete, under \tt{FP} reductions, and let $B\in\tt{mNP}$.
From Theorem \ref{thm:logical_characterization}, let $F'\in\tt{posFP}$ such that $x\in A$ iff there is $y\leq x$ and $w\in\{0,1\}^{q(|x|)}$ such that $F'(\diamond y,\diamond w) = 1$.
Let $\{C_{n+m}\}$ be the circuit family computing $F'$.
Then, from Theorem \ref{thm:diamond}, the function $x\mapsto \diamond\code{C_{2(|x|+q(|x|))}}$ is in \tt{posFP}.
Take a \textit{predictable} \tt{FP} reduction $G$, from \tt{CSAT} to $\Xi(A)$, and let $G'\in\tt{posFP}$ simulate $G$ in the sense of Theorem \ref{thm:diamond}.
Then, $x\mapsto \diamond\code{C_{2(|x|+q(|x|))}} \mapsto G'(\diamond\code{C_{2(|x|+q(|x|))}})$ is \textit{almost} a reduction from $B$ to $A$.
We only have to further impose the condition given by $\chi_\leq$ in the instance constructed. 

Unfortunately, this inequality depends on the precise value of $x$, so we cannot construct its negation using a \tt{posFP} function.
\end{rmrk}

We give in the preceding sections two examples where the function $z\mapsto \chi_\leq(z,x)$ can be internalized in a reduction using only positive functions.
\subsection{Monotone circuit satisfiability}\label{sec:mCSAT}
The set \tt{mCSAT} contains the pairs $\pair{\code{C},x}$ for which there is $z\in \{0,1\}^{|x|}$ such that $C(\diamond z) = 1$. Formally:

\begin{defn}\label{def:mCSAT}
$\tt{mCSAT} := \{x\in\bb{W}\,|\,\exists z\in\{0,1\}^{|(x)_1|}:\tt{EVAL}((x)_0,\diamond z)=1\}$.
\end{defn}

To prove its completeness, we will follow the characterization of \tt{mNP} given by Theorem \ref{thm:logical_characterization}. As it requires the use of the function $\chi_\leq$, we first need to show that this function can be constructed by positive functions. Note that this function requires the precise value of $x$.

\begin{lemm}\label{lemm:leq_circuit}
There is a function in \tt{posFP} which, given $x$, returns the code of the circuit computing $z\mapsto\chi_\leq(z,x)$, for inputs $z\in\{0,1\}^{2|x|}$. 
\end{lemm}
\begin{proof}
Given the construction of $\chi_\leq$ from Lemma \ref{lemm:leq_diamond}, we need to construct a \tt{posFP} function which, given $x$, returns the code of the circuit $C_{2|x|}$ computing $\bigwedge_{i=0}^{|x|-1}(z(2i)\vee x(i))$.
Let $|x| = n$.
Since the circuit $C_{2n}$ has $2n$ inputs, it has $2n+2$ initial gates, one for each input bit and $2$ for the constants $0$ and $1$.

First, consider, for each $0\leq i \leq n-1$, the \tt{posFP} function $g_i$, which given $x$ returns
\begin{align*}
    &\underbrace{10}_{\vee\text{ gate}}\oplus\,\, 0^{2n-2i-2}\oplus \underbrace{01}_{z(2i)}\oplus \,\,0^{2i}\oplus \underbrace{0\oplus x(i)}_{\text{Connection to the constants}}\oplus \underbrace{0^{i}}_{\text{no further connections}}&
\end{align*}
The initial $10$ indicates a $\vee$ gate. The word $0^{2n-2i-2}\oplus 01 \oplus 0^{2i}$, of size $2n$, indicates that this gate is only connected to the input gate corresponding to $z(2i)$ (when $z=\diamond w$, this is the bit that is negated). The bits $0x(i)$ indicate no connection to the $0$ constant, and a connection to the $1$ constant iff $x(i) = 1$. The word $0^{i}$ indicates that the gate is not connected to any of the previous gates.

Thus, with only positive operations, we may use the value of $x(i)$ to represent the presence or absence of an edge to the constant $1$. For example, if $|x|=2$, the first gate will have the code $10\oplus 00\oplus 01\oplus 0x(0)$ (representing $z(0)\vee x(0)$) and the second the code $10\oplus01\oplus 00\oplus 0x(1)\oplus 0$ (representing $z(2)\vee x(1)$).

Concatenating, for each $0\leq i \leq n-1$, the value of $g_i(x)$, one obtains the expressions $z(2i)\vee x(i)$: for the gate to return $1$, either $z(2i)=1$, or $x(i)=1$.
Now, we must merge these gates with a $\wedge$ gate. 
Let $\tt{AND}$ be the function which, given $x$, returns
\begin{align*}
    &\underbrace{01}_{\wedge\text{ gate}}\oplus\underbrace{1^{2n+2}}_{\text{The input and constants}}\oplus \underbrace{0^n}_{\text{The previous gates}}&
\end{align*}

Thus, the following \tt{posFP} function returns the code of the desired circuit:
\begin{align*}
    &x\mapsto g_0(x)\oplus g_1(x)\oplus\cdots\oplus g_{n-1}(x)\oplus \tt{AND}(x)&
\end{align*}
\end{proof}

\begin{thm}\label{thm:mCSAT_mNP_completeness}
\tt{mCSAT} is \tt{mNP} complete under \tt{posFP} many-to-one reductions.
\end{thm}
\begin{proof}
Since \tt{mCSAT} is a monotone problem and is in \tt{NP}, it is in $\tt{mNP} = \tt{NP}\cap\tt{mono}$.

Now, let $A\in\tt{mNP}$. From the proof of Theorem \ref{thm:logical_characterization} there is a monotone polynomial $q$ and $F\in\tt{posP}$ such that $x\in A$ iff there is $y\leq x$ and $z\in\{0,1\}^{q(|x|)}$ such that $F(\diamond y,\diamond z) = 1$.

Let $\{C_{n+m}\}$ be the \tt{posP}-uniform monotone circuit family computing $F$.
Then, $x\in A$ iff there is $y\leq x$ and $z\in\{0,1\}^{q(|x|)}$ such that $C_{2(|x|+q(|x|))}(\diamond y\diamond z) = 1$.

For each $x\in\bb{W}$, let $\{L_n^x\}_{n\geq1}$ be the \tt{posP}-uniform monotone circuit family computing $z\mapsto \chi_\leq(z(|z|-1)\cdots z(|z|-2|x|,x)=1$, for inputs in $\{0,1\}^n$ (the circuit is analogous to the one from Lemma \ref{lemm:leq_circuit}, but it ignores the remaining bits of $z$).

Then, $x\in A$ iff there is $z\in\{0,1\}^{|x|+q(|x|)}$ such that $L_{2|z|}^x(\diamond z) \wedge C_{2|z|}(\diamond z) = 1$.
Denote the conjunction of these circuits by $E_n^x$. Then, the function $(x,n)\mapsto \code{E_n^x}$ is in \tt{posFP}, whence the function $x\mapsto \pair{\code{E_{2(|x|+q(|x|))}^x},x\oplus 1^{q(|x|)}}$ is a \tt{posFP} reduction from $A$ to \tt{mCSAT}.
\end{proof}

From the separation of \tt{posP} from \tt{mNP}, it follows:

\begin{thm}\label{thm:mcsat_notin_posP}
$\tt{mCSAT}\notin\tt{posP}$.
\end{thm}

Note that, alternatively to the construction above, one could proceed as follows. 
First, construct the (code of the) circuits $L_n^{|x|}$, similarly to Lemma \ref{lemm:leq_circuit}, but for the function $z\mapsto \chi_\leq(z,1^{|x|})$ instead of $z\mapsto \chi_\leq(z,x)$. 
Then, given the code of $E_n^{|x|}$, which, in this case, would depend only on $|x|$, one would only have to \textit{intersect} each bit $x(i)$ with the bit $1$ in the position establishing a connection of the $\vee$ gate with the constant $1$.

The outcome would be the same, but clearly separates the construction of an instance of \tt{mCSAT}, which only depends on $|x|$, with the postprocessing, which simply requires an intersection of parts of the instance with the concrete value of $x$.
The reduction we show in the next section follows these lines as well.
\subsection{The Hamiltonian cycle problem}
We prove the \tt{mNP}-completeness of the Hamiltonian cycle problem (\tt{HCP}) using the \tt{mNP}-completeness of \tt{mCSAT}, Theorem \ref{thm:diamond}, and a composition of known reductions.

\tt{HCP} consists of the sets of graphs for which there is a cycle passing through each vertex only once (a Hamiltonian cycle). 
This is clearly a \textit{monotone} problem: if an edge is added to a graph, then the presence of a Hamiltonian cycle is preserved.

\begin{defn}\label{def:paths}
For $x$ of size $n^2$, denote by $x[i,j]$ the bit in the $ni + j$th position of $x$.
Then, $x\in\{0,1\}^{n^2}$ encodes the graph with $n$ vertices and an oriented edge from $i$ to $j$, whenever $x[i,j] = 1$.
\end{defn}

For a formulation of \tt{HCP} in the style of Theorem \ref{thm:logical_characterization}, let paths be encoded as adjacency matrices as well, where each row contains only one $1$.
Then, due to Theorem \ref{thm:diamond}, there is a \tt{posP} function \tt{CYCLE} which, given $\diamond p$ and a graph $x$, can test if $p$ represents a cycle of $x$ and if $|x|$ is a perfect square.
Checking if $p$ represents a cycle is simple, and, if $\chi_\leq(\diamond p,x) = 1$, then $p\leq x$, so each edge of the cycle represented in $p$ also exists in the graph $x$. 
Thus, one may formulate \tt{HCP} as follows:

\begin{defn}\label{def:HCP}
$\tt{HCP} = \{x\in\bb{W}\,|\,\exists z\in\{0,1\}^{|x|}\,:\,\tt{CYCLE}(x,\diamond z)=1\}$.
\end{defn}

\begin{thm}\label{thm:hamiltonian_mNP_complete}
The Hamiltonian cycle problem is \tt{mNP} complete.
\end{thm}
\begin{proof}
Let $A\in\tt{mNP}$. We proceed by composing known reduction, ensuring that they are \textit{predictable}.

\smallskip
\textbf{From $A$ to a circuit}.
From the proof of Theorem \ref{thm:mCSAT_mNP_completeness}, let $\{C_n\}$ be a \tt{posP}-uniform monotone circuit family such that $x\in A$ iff there is $z\in\{0,1\}^{|x|+q(|x|)}$ such that $\chi_\leq(z(|z|-1)\cdots z(|z|-2|x|),x)=1$ and $C_{2|z|}(\diamond z) = 1$. 
Note that $C_{2|z|}$ depends only on the size of $x$ (the function $\chi_\leq$ is not yet internalized).
Thus, from Theorem \ref{thm:diamond}, $x\mapsto \diamond \code{C_{2(|x|+q(|x|))}}$ is in \tt{posFP}.

\smallskip
\textbf{From a circuit to a CNF formula}.
We apply Tseitin's transform \cite{tseitin:83} (see e.g. \cite[Chapter $22$]{sannella:22}). 
Suppose a circuit has $2n$ input gates and $k$ non-input gates.
The input gates are assigned the variable $v_{1},\dots,v_{2n}$ and the remaining gates the variables $v_{2n+1},\dots,v_{2n+k}$. 
For each gate $2n+1\leq i\leq 2n+k$, we construct:
\begin{itemize}
	\item[] if $v_i = v_j\wedge v_l$, for $j,l<i$, the expression $(\neg v_i \vee v_j) \wedge (\neg v_i \vee v_l) \wedge (v_i \vee \neg v_j\vee \neg v_l)$
	\item[] if $v_i = v_j\vee v_l$, for $j,l<i$, the expression $(v_i \vee \neg v_j) \wedge (v_i \vee \neg v_l) \wedge (\neg v_i \vee v_j\vee v_l)$
\end{itemize}
The conjunction of all these clauses gives a \tt{CNF} formula $\phi$ that is satisfiable iff $C$ is satisfiable.
We must still enforce that the variables $v$ are in the form $\diamond w$, so we construct the formula
\begin{align*}
	&\phi' = \phi\,\wedge\,\bigwedge_{0\leq i \leq n-1}(\neg v_{2i+1}\vee \neg v_{2i+2})\wedge ( v_{2i+1}\vee v_{2i+2})&
\end{align*}
Then, $\phi'$ is satisfiable iff there is $z\in \{0,1\}^{n}$ such that $C(\diamond z) = 1$.
Moreover, $\phi'$ has $2n + k$ variables (one added for each gate) and $3k + 2n$ clauses ($3$ for each $\wedge$ and $\vee$ gate, plus $2n$ to enforce the $\diamond$ structure).
This size depends only on the number of gates of $C$, so, from Theorem \ref{thm:diamond} the function $x\mapsto \diamond\code{C}\mapsto\diamond\code{\phi'}$ is in \tt{posFP}. 

\smallskip
\textbf{From a CNF formula to a Graph}.
For a reduction from $\tt{CNF-SAT}$ to \tt{HCP}, we take an example similar to the one given in \cite[Chapter $8$]{kleinberg:05}. As we require its specific form, we start by describing it.\footnote{A good visualization of this reduction is given in Chapter $11.18$ of \url{https://chalmersgu-data-structure-courses.github.io/OpenDSA/Published/ChalmersGU-DSABook}.}
If $\phi$ has $n$ variables and $k$ clauses, one constructs a graph with: $k$ vertices, one per each clause, $C_1,\dots,C_k$, plus $n\cdot (2k +2)$ vertices, one for each variable, plus $n+1$ ``middle vertices''.

For example, if a formula has two clauses, $C_1$ and $C_2$, to each variable $v$ would correspond a block as:
{\small
\begin{center}
\begin{tikzpicture}[node distance=1.5cm, auto]
    \node[draw, circle] (x) {$v$};
    \node[above right of=x] (sx) {$\cdots$};
    \node[below right of=x] (tx) {$\cdots$};
    \node[draw, circle, right of=x] (1) {};
    \node[draw, circle, right of=1] (2) {};
    \node[draw, circle, right of=2] (3) {};
    \node[draw, circle, right of=3] (4) {};
    \node[draw, circle, right of=4] (5) {$v$};
    \node[above left of=5] (s5) {$\cdots$};
    \node[below left of=5] (t5) {$\cdots$};

    \draw[->, thick] (sx) to node {} (x);
    \draw[->, thick] (x) to node {} (tx);
    \draw[->, thick] (s5) to node {} (5);
    \draw[->, thick] (5) to node {} (t5);

    \draw[->, thick] (x) to [bend left] node {} (1);
    \draw[->, thick] (1) to [bend left] node {} (2);
    \draw[->, thick] (2) to [bend left] node {} (3);
    \draw[->, thick] (3) to [bend left] node {} (4);
    \draw[->, thick] (4) to [bend left] node {} (5);

    \draw[<-, thick] (x) to [bend right] node {} (1);
    \draw[<-, thick] (1) to [bend right] node {} (2);
    \draw[<-, thick] (2) to [bend right] node {} (3);
    \draw[<-, thick] (3) to [bend right] node {} (4);
    \draw[<-, thick] (4) to [bend right] node {} (5);

    \node[draw, circle, above of=2, node distance=1.5cm] (C1) {{\tiny $C_1$}};
    \node[draw, circle, above of=3, node distance=1.5cm] (C2) {{\tiny $C_2$}};

    \draw[->, thick] (1) -- (C1);
    \draw[->, thick] (C1) -- (2);

    \draw[->, thick] (4) -- (C2);
    \draw[->, thick] (C2) -- (3);
\end{tikzpicture}
\end{center}
}

Where a connection to $C_1$ from left to right indicates that $v$ appears in $C_1$ positively, and a connection to $C_2$ from right to left indicates that $v$ appears in $C_2$ negatively.
In this example, a Hamiltonian path which traverses this block from right to left indicates that $v = 0$, so the clause $C_2$ is satisfied by $v$.

Now, both outer vertices of each block (one for each variable) are connected to an extra middle vertex, which is in turn connected to both outer edges of the next block (these are $n-1$ of the ``middle'' vertices).
Finally, a starting vertex $s$ and a terminal vertex $t$ are added (the remaining ``middle vertices''), with $s$ connected to the edges of the first block, $t$ connected to $s$, and the edges of the last block connected to $t$.

Overall, the graph has then $n\cdot(2k +2) + k + n+1$ vertices and a Hamiltonian cycle iff $\phi$ is satisfiable.
Thus, from Theorem \ref{thm:diamond}, the function $x\mapsto\diamond\code{C}\mapsto\diamond\code{\phi'}\mapsto\code{G}$, returning the code of the described graph, is in \tt{posFP}, and almost our desired reduction.

\smallskip

The function $x\mapsto\code{G}$ gives a graph that has a Hamiltonian cycle iff there is $z\in\{0,1\}^{|x|+q(|x|)}$ such that $C_{2(|x|+q(|x|))}(\diamond z) = 1$. However, the inequality with $x$ still has to be imposed.

That is, we must change the graph so that it has a Hamiltonian cycle iff $\phi'$ is satisfiable \textit{and} $\chi_\leq(v_1\cdots v_{2|x|},x) = 1$, i.e, $v_1v_3\cdots v_{2|x|-1}\leq x$.
This inequality is represented by $v_{2i-1}\leq x(|x|-i)$, for all $1\leq i\leq |x|$.
To enforce it, we simply intersect $x(|x|-i)$ with the part of the adjacency matrix of $G$ that represents the left to right cycle of the variable $v_{2i-1}$.
If $x(|x|-i) = 0$ those paths are removed, so one obtains:

{\small
\begin{center}
\begin{tikzpicture}[node distance=1.5cm, auto]
    \node[draw, circle] (x) {{\tiny $v_{2i-1}$}};
    \node[above right of=x] (sx) {$\cdots$};
    \node[below right of=x] (tx) {$\cdots$};
    \node[draw, circle, right of=x] (1) {};
    \node[draw, circle, right of=1] (2) {};
    \node[draw, circle, right of=2] (3) {};
    \node[draw, circle, right of=3] (4) {};
    \node[draw, circle, right of=4] (5) {{\tiny $v_{2i-1}$}};
    \node[above left of=5] (s5) {$\cdots$};
    \node[below left of=5] (t5) {$\cdots$};

    \draw[->, thick] (sx) to node {} (x);
    \draw[->, thick] (x) to node {} (tx);
    \draw[->, thick] (s5) to node {} (5);
    \draw[->, thick] (5) to node {} (t5);

    \draw[->, thick] (1) to node {} (x);
    \draw[->, thick] (2) to node {} (1);
    \draw[->, thick] (3) to node {} (2);
    \draw[->, thick] (4) to node {} (3);
    \draw[->, thick] (5) to node {} (4);

    \node[draw, circle, above of=2, node distance=1.5cm] (C1) {{\tiny $C_1$}};
    \node[draw, circle, above of=3, node distance=1.5cm] (C2) {{\tiny $C_2$}};

    \draw[->, thick] (1) -- (C1);
    \draw[->, thick] (C1) -- (2);

    \draw[->, thick] (4) -- (C2);
    \draw[->, thick] (C2) -- (3);
\end{tikzpicture}
\end{center}
}

This ensure that if a cycle through $G$ exists, it must take the right to left direction on the block of $v_{2i-1}$, representing an assignment of $\phi$ where $v_{2i-1} = 0$.
Repeating this operation for the variables $v_{1},v_{3},\cdots,v_{2|x|-1}$, the desired inequality is imposed. Moreover, only positive functions are required.

We thus obtain a \tt{posFP} reduction from $A$ to \tt{HCP}, which concludes the proof.
\end{proof}

Note that, besides the initial use of Theorem \ref{thm:diamond}, to construct a circuit and its negation, and the final embedding of $\chi_\leq(\,\cdot\,,x)$ in the graph, the middle steps are simply an analysis of \textit{known} reductions, to ensure that they are \textit{predictable}.
\section{Contradictory relativizations}
We extend the previous approach to positive relativized computation and prove the \tt{mPSPACE} completeness of an extension of \tt{mCSAT}.
From here, a contradictory relativization for the separation $\tt{posP}\subsetneq\tt{mNP}$ will be obtained.
\subsection{Relativized positive machines}\label{sec:relativized_TM}
\begin{defn}\label{def:oracle_turing machine}
An oracle Turing machine is a Turing machine with an extra tape, the \textit{query} tape and three extra states: $q_{query}$, $q_{yes}$ and $q_{no}$.
An oracle Turing machine has an oracle for $A$ if, whenever it enters the query state $q_{query}$ with $w\in\bb{W}$ written on its query tape, the next state is $q_{yes}$, if $w\in A$, and $q_{no}$, otherwise.
\end{defn}

We now have to extend the ordering on configurations (see Definition \ref{def:order}) to the oracle states. 

\begin{defn}\label{def:positive_oracle_turing_machine}
For $M$ an oracle Turing machine, with oracle states $q_{query}$, $q_{yes}$ and $q_{no}$, define the partial orders $\preceq_\frak{I}$ and $\preceq_\frak{O}$ as follows: 
\begin{itemize}
    \item[] $(q,\sigma_1,\cdots,\sigma_k)\preceq_\frak{I} (q',\sigma'_1,\cdots,\sigma'_k)$ if: either $q=q'$, or $q=q_{no}$ and $q'=q_{yes}$, and, for $1\leq j\leq k$, either $\sigma_j =\sigma'_j$, or $\sigma_j =0$ and $\sigma'_j=1$
    \item[] $(q,\tau_1, d_1,\cdots,\tau_k,d_k)\preceq_\frak{O} (q',\tau'_1, d'_1,\cdots,\tau'_k,d'_k)$ if: either $q=q'$, or $q=q_{no}$ and $q'=q_{yes}$, and, for $1\leq j\leq k$, $d_j=d'_j$ and either $\tau_j =\tau'_j$, or both $\tau_j =0$ and $\tau'_j=1$
\end{itemize} 
\end{defn}

Essentially, the oracle states $q_{no}$ and $q_{yes}$ are comparable to each other (with $q_{no}$ smaller than $q_{yes}$), and all remaining states are incomparable as before.
This definition of monotonicity ensures the following property: 

\begin{rmrk}\label{rmrk:monotone_oracle_relativized}
If $M$ is a positive oracle Turing machine machine and $A\subseteq\bb{W}$ is a monotone set, then the function computed by $M^A$ is monotone.   
\end{rmrk}

See the Appendix \ref{sec:oracle_order} for an example where this property would break, if a different ordering was considered.

\begin{defn}\label{def:turing_completeness}
$A$ is \textit{complete} for $\cal{C}$ under \tt{posP} Turing reductions if:
for every $B\in\cal{C}$, there is an polynomial time oracle posDTM $M$ such that $x\in B$ iff $M^A(x)=1$. We abbreviate the latter condition as $B\leq_T^{pos} A$.   
\end{defn}

As a consequence of this definition, the following simple observation follows:

\begin{prop}\label{prop:complete_for_P}
The set $A=\{0\}$ is complete for \tt{P} under \tt{posP} Turing reductions.
\end{prop}
\begin{proof}
Let $B\in\tt{P}$.
Define the $\tt{posP}^A$ machine $M'$ as follows:
on input $x$, query $A$ with each bit $x(i)$. 
If $x(i)\in A$, append to the work tape $x(i)1$;
if $x(i)\notin A$, append to the work tape $x(i)0$.
After the bits of $x$ have all been read, the work tape contains the word $\diamond x$.
Then, from Theorem \ref{thm:diamond}, $1_B(x)$ can be simulated by a positive machine on input $\diamond x$.
\end{proof}

One can show that $\{0\}$ is not complete for \tt{P} under \tt{posFP} many-to-one reductions, so in general, many-to-one and Turing positive reductions are \textit{not} equivalent.
Of course, completeness under the first implies completeness under the latter.
\subsection{Contradictory relativization}
We reuse the techniques from Section \ref{sec:positive_many-to-one_completeness} and construct a problem that is complete for \tt{mPSPACE}.
This will be a generalization of \tt{mCSAT} by allowing alternation \cite{kozen:81} in the acceptance criteria.
Abbreviate $Q_0$ and $Q_1$ for the quantifiers $\forall$ and $\exists$, restricted over $\{0,1\}$.
For example, the word $001$ represents a quantification of the form $\forall x\forall y\exists z$. 

\begin{defn}\label{def:mTQC}\,
\begin{align*}
    &\tt{mTQC} := \{x\in\bb{W}\,|\,\,Q_{(x)_1(|(x)_1|-1)}v_1\cdots Q_{(x)_1(0)}v_{|(x)_1|}\,,\,
    \tt{EVAL}((x)_0,\diamond v_1\oplus\cdots\oplus v_{|(x)_1|})=1\}&
\end{align*}
\end{defn}

This is monotone version of the circuit version of the problem True Quantified Boolean Formula (\tt{TQBF}), which is \tt{PSPACE}-complete \cite{stockmeyer:73}.
For example, $\pair{\code{C},101}\in\tt{mTQC}$ iff 
\begin{align*}
    &\exists v_1\in\{0,1\}\forall v_2\in\{0,1\}\exists v_3\in\{0,1\}\,:\,C(\diamond v_1v_2v_3) = 1&
\end{align*}
The set \tt{mCSAT} can be viewed as a particular case of \tt{mTQC} where $(x)_1\in\{1\}^*$, representing thus only existential quantifiers.
Analogously to \tt{mCSAT}, one may show that \tt{mTQC} is a monotone problem (if a $0$ becomes a $1$, in the word representing the quantification, a universal quantifier becomes an existential quantifier).
Since it is in \tt{PSPACE}, it is also in \tt{mPSPACE}, which is a trivial class \cite{grigni:92}.

Since positive many-to-one completeness implies positive Turing completeness, we prove the first, even though we will use the latter.

\begin{thm}\label{thm:mTQC_mPSPACE_complete}
\tt{mTQC} is \tt{mPSPACE}-complete under \tt{posFP} many-to-one reductions.    
\end{thm}
\begin{proof}
The proof is analogous to the proof of Theorem \ref{thm:mCSAT_mNP_completeness}, so we only sketch it.
Suppose that $A\in\tt{mPSPACE}$.
Since $\tt{PSPACE} = \tt{AP}$ \cite{kozen:81}, there is an alternating Turing machine $M$ deciding $A$ in polynomial time.
We may assume that there is some monotone polynomial $q$ such that $M$ starts its computation on $x$ by alternating $q(|x|)$ times between existential and universal quantifiers, each time guessing $q(|x|)$ bits, and then runs a polynomial time function on $x$ and the guessed bits $z$, $F(x,z)$.

Since $A$ is monotone, it is also decidable by the alternating Turing machine $M'$ which, on input $x$, starts by guessing $y\leq x$ (with existential states) and then operates as $M$ on input $y$, rather than on $x$.
Using Theorem \ref{thm:diamond}, we may thus assume that $M'$ starts by guessing $y\leq x$, then alternates between existential and universal states $q(|x|)$ times, each time guessing $q(|x|)$ bits, and then computes $F'(\diamond y,\diamond z) = F(y,z)$, for $F'\in\tt{posP}$.

This alternation mechanism is represented, for each $x\in\bb{W}$, by the word $k(x) = 1^{|x|}1^{q(|x|)}0^{q(|x|)}\cdots$ of size $|x|+q(|x|)^2$.
Then, just as in Theorem \ref{thm:mCSAT_mNP_completeness}, the \tt{posP} functions $F'$ and $\chi_\leq(\,\cdot\,,x)$ can be encoded in a circuit family $\{E_n^x\}$ so that the function $x\mapsto \pair{\code{E_{2(|x|+q(|x|)^2)}^x},k(x)}$ is a \tt{posFP} reduction from $A$ to \tt{mTQC}
\end{proof}

\begin{thm}\label{thm:contradictory_relatvization}
There is a monotone oracle $A$ such that $\tt{posP}^A=\tt{mNP}^A$.
\end{thm}
\begin{proof}
Let $A:=\tt{mTQC}$.
Since sets in $\tt{mPSPACE}^A$ are monotone (as $A$ is monotone) and at most in $\tt{PSPACE}^\tt{PSPACE} = \tt{PSPACE}$, sets in $\tt{mPSPACE}^A$ are in $\tt{mPSPACE}$.

Since $A$ is \tt{mPSPACE} complete under \tt{posFP} reductions, $\tt{mPSPACE}\subseteq \tt{posP}^A$, as the \tt{posP} function can simply implement the reduction and query $A$.

Since $\tt{posP}\subseteq \tt{mNP}\subseteq \tt{mPSPACE}$, and these inclusion relativize, we have that $\tt{posP}^A\subseteq\tt{mNP}^A\subseteq \tt{mPSPACE}^A$.

Altogether, this gives
\begin{align*}
    &\tt{mPSPACE}\subseteq \tt{posP}^A\subseteq \tt{mNP}^A \subseteq  \tt{mPSPACE}^A = \tt{mPSPACE}&
\end{align*}
Thus, we obtain that $\tt{posP}^A = \tt{mNP}^A$.
\end{proof}
\section{A weak diagonalization proof that \texorpdfstring{$\tt{posP}\neq\tt{mNP}$}{pos P neq mNP}}\label{sec:diagonalization}
Before we exhibit a diagonalization proof that $\tt{posP}\neq\tt{mNP}$, we briefly discuss the ideas behind \cite{kozen:78} and \cite{nash:03}.
Kozen \cite{kozen:78} showed that, if $\tt{P}\neq\tt{NP}$, there is a proof that $\tt{P}\neq\tt{NP}$ by diagonalization. However, the notion of diagonalization he considered is rather weak, requiring one to simply prove that there is some computable enumeration of \tt{P} whose \textit{diagonal} is in \tt{NP}.
Usually, one would firsts show that the \textit{enumeration} of $\tt{P}$ is in \tt{NP}, and then from this fact conclude that so is its diagonal (the last step uses the closure of \tt{NP} under linear time many-to-one reductions).

Nash et. al. \cite{nash:03} termed a diagonalization in the style of Kozen a \textit{weak diagonalization}, and a diagonalization satisfying the latter condition a \textit{strong diagonalization} .

We present here a \textit{weak diagonalization} proof that $\tt{posP} \subsetneq \tt{mNP}$.
Besides the lemma below, which is particular to monotone computation, the construction is a simple adaptation of \cite{kozen:78}.

\begin{lemm}\label{lemm:fin_var}
Suppose that $F\notin\tt{posP}$ is \textit{monotone}. Then, $F$ is not a finite variant of any \tt{posP} function. 
\end{lemm}
\begin{proof}
Suppose that $F$ was a finite variant of some function $f\in\tt{posP}$.
For simplicity, assume that they only differ in $\tilde{x}\in\{0,1\}^n$.

Since $F$ is monotone, if $F$ accepts $x\in\{0,1\}^n$, it accepts all $y\geq x$.
Thus, for each $x\in\{0,1\}^n$, let $c_x$ be the smallest word (comparable with $x$) in $\{0,1\}^n$ such that $F(c_x)=1$. 
Let $b_n$ be the number of distinct words meeting the previous condition. Then, we may enumerate them as $k_1^n,\dots,k_{b_n}^n$ and, for all $x\in\{0,1\}^n$, $F(x)=\bigvee_{j=1}^{b_n} \chi_\leq(\diamond k_j^{n},x)$.

Then, we construct the \tt{posP} function $f'$ as follows:
\begin{align*}
    &f'(x) = 
    \begin{cases}
        f(x) & |x|\neq n\\
        \bigvee_{j=1}^{b_n} \chi_\leq(\diamond k_j^{n},x) & |x| = n
    \end{cases}&
\end{align*}
Since $f$ and $\chi_\leq$ are positive, and the constants $\diamond k_j^{n}$ don't depend on the input, $f'\in\tt{posP}$. Moreover, $f' = F$, which contradicts $F\notin\tt{posP}$.
\end{proof}

Note that the case distinction in the proof cannot be made simply on $\tilde{x}$, as this would require negative operations to compare the input with $\tilde{x}$. Rather, all inputs of size $|\tilde{x}|$ must be ``corrected''.
Moreover, this result is not true for arbitrary (non-monotone) functions. For example, $1_{\{0\}}$, the characteristic function of the set $\{0\}$, is not in \tt{posP}, and yet it is a finite variant from the constantly $0$ function.

\begin{defn}\label{def:diagonal}
Given an enumeration $U$ of a complexity class $\cal{C}$, the main diagonal of $U$ is the function $diag_U$ such that
\begin{align*}
    &diag_U(x) = 
    \begin{cases}
        1 & \text{ if } U(x,x) = 0\\
        0 & \text{ otherwise} 
    \end{cases}&
\end{align*}
\end{defn}  

From now on, fix some (not necessarily monotone) universal function $U$ of \tt{posP}. 
Since, for every $x\in\bb{W}$, $diag_U(x)\neq U(x,x)$, $diag_U\notin\tt{posP}$.

Abusing notation, let $\tt{HCP}$ also denote the \textit{monotone} function deciding \tt{HCP}. We give an enumeration of \tt{posP} whose main diagonal is \tt{HCP}.

\begin{thm}\label{thm:diagonalization}
There is an enumeration of \tt{posP} whose diagonal is $\tt{HCP}$.
\end{thm}
\begin{proof}
For $i\in\{0,1\}$, let $const(i)$ be a code of the constant function $x\mapsto i$ in the enumeration $U$.
Define
\begin{align*}
    &f(x) = 
    \begin{cases}
        \mu y\,:\,\tt{HCP}(y) = 1 - U(\eps,y) \text{ if } x=\eps\\
        \mu y\,:\,\big(y>f(x-1)\,\wedge\,\,\tt{HCP}(y) = 1 - U(x,y)\big) & \text{otherwise}
    \end{cases}&
\end{align*}
which, given an index $x$, returns the first point $y$ where $\tt{HCP}(y)\neq U(x,y)$.
Since $\tt{HCP}$ is not a finite variant of any \tt{posP} function (Lemma \ref{lemm:fin_var}), $f$ is well defined and, for each $x$, $U(x,f(x)) = 1 - \tt{HCP}(f(x))$.

Now, we define a ``translation function'' $h$, which will convert codes from the enumeration $U$ to a new enumeration $V$. Define $h$ as
\begin{align*}
    h(x)&=
    \begin{cases}
        f^{-1}(x)   &   \text{if  }  x\in ran(f)\\
        const(0)    &   \text{if  }  \tt{HCP}(x) = 1\\
        const(1)    &   \text{otherwise}
    \end{cases}
    &
\end{align*}

As $f$ is increasing and computable, the range of $f$ is decidable, whence $h$ is computable as well. 
Let $V(x,y):= U(h(x),y)$. This function is the desired enumeration.

Since, for every $x\in\bb{W}$, $h(f(x)) = f^{-1}(f(x)) = x$, $h$ is onto, so $V$ is also an enumeration of \tt{posP}.
Now, we show that $diag_V  = \tt{HCP}$.

Let $x\in\bb{W}$.
If $x\notin ran(f)$, then from the construction of $h$,
\begin{align*}
    V(x,x) &= U(h(x),x) = 
    \begin{cases}
        U(const(0),(x))  &   \text{if  } \tt{HCP}(x) = 1\\
        U(const(1),(x))   &   \text{otherwise}
    \end{cases}
    = 1- \tt{HCP}(x)&
\end{align*}
If $x = f(y)\in ran(f)$, then $h(x) = f^{-1}(x) = y$, so
\begin{align*}
    V(x,x)&= U(h(x),x) = U(y,f(y)) = 1 - \tt{HCP}(f(y)) = 1 - \tt{HCP}(x)&
\end{align*}
Thus, in both cases, $V(x,x) = 1 - \tt{HCP}(x)$, which concludes the proof.
\end{proof}

The above is a \textit{weak} diagonalization proof that $\tt{posP}\neq\tt{mNP}$, as nothing is proven about the enumeration $V$, but rather only about its diagonal.
\section{Conclusion}
In this paper we explored completeness of \textit{monotone} problems under \tt{posFP} reductions and relativization of positive models of computation.
While we focused on the classes \tt{mNP} and \tt{mPSPACE}, there are also interesting questions below \tt{mNP}. 
For example, are there sets in \tt{mNP} that are not complete under \tt{posP} reductions (recall \cite{ladner:75,balcazar:82}). 
And does \tt{mP} have complete problems, under \tt{posFP} reductions?

Regarding \tt{mNP}, the first observation, also made in \cite{skapinakis:26}, is that \tt{mNP} problems can be written in the form $\exists z\leq q(|x|):B(x,\diamond z)$, for $B$ \textit{positive}. This suggests a quantifier characterization of \tt{mNP} and related monotone non-deterministic classes in the style of \cite{wrathall:76,toran:91}, or even \cite{buss:85,strahm:03}.

From Section \ref{sec:positive_many-to-one_completeness}, two natural monotone problems that are \tt{NP}-complete are also \tt{mNP} complete under \tt{posFP} many-to-one reductions. The question of whether this is true for \textit{every} monotone \tt{NP}-complete problem remains open, and is a sufficient condition for $\tt{P}\neq\tt{NP}$.
One of the challenges in constructing positive reductions is the encoding  of a condition of the form $y\leq x$. 
In the case of \tt{mCSAT}, this restriction corresponds to \textit{not} adding certain connections in a $\vee$ gate to the constant $1$;
in the case of \tt{HCP}, it corresponds to removing certain edges from a graph.

Note that there is an interesting duality in this condition: if positive reductions are \textit{strong} enough to simulate reductions on monotone sets, then \tt{P} is \textit{weaker} than \tt{NP}.
There is another way to see this implication. If $\tt{P}=\tt{NP}$, then all \tt{P} problems are \tt{NP}-complete under \tt{FP} reductions, as the reduction can simply itself solve the problem. Thus, $\tt{P}=\tt{NP}$ implies the existence of reductions that are too complicated to be simulated by \tt{posFP} functions.

With regards to relativized models of computation, in Section \ref{sec:relativized_TM} we introduced positive oracle Turing machines, which compute monotone functions whenever they are relativized to a monotone oracle.
Based on the completeness of \tt{mTQC} for \tt{mPSPACE}, it follows that there is a \textit{monotone} oracle for which \tt{posP} equals \tt{mNP}.

This result is motivated by a discussion which dates back to Baker, Gill and Solovay \cite{baker:75}, about the role of relativization in complexity theory and the consequence of contradictory relativizations as \textit{complexity barriers} \cite{fortnow:94,fortnow:00,nash:03,aaronson:06}.
What \textit{precisely} makes a proof technique yield a non-relativizing result is not clear \cite{arora:91,fortnow:94,impagliazzo:09}, and neither is what counts as a \textit{genuine} non-relativizing result \cite[Section $4$]{fortnow:94}.
We have thus contributed with a new (old) example of a non-relativizing separation, for which a (weak) diagonalization can even be constructed.
The interesting point about that proof is that it shows (as has been remarked in e.g. \cite{nash:03}) that \textit{diagonalization} is not necessarily a relativizing proof technique. In our case, it is \textit{precisely} the fact that the enumeration $V$ \textit{is indeed} an enumeration which does not relativize -- there is an oracle for which the \textit{translation function} $h$ is not defined.

Finally, this paper was motivated by recent work on \textit{implicit} characterizations of monotone classes \cite{das:16,das:18,skapinakis:26}.
Indeed, it is possible to adapt the theories given in \cite{strahm:03} to give proof-theoretic characterizations of \textit{positive} complexity classes that are known to be different.
It is the goal of current research to study these separations from a logical perspective and to develop new proof-theoretic based methods to prove complexity lower bounds.
\section*{Acknowledgements}
This work was funded by national funds through the FCT – Fundação para a Ciência e a Tecnologia, I.P., under the scope of the projects UIDB/00297/2020 (\url{https://doi.org/10.54499/UIDB/00297/2020}) and UIDP/00297/2020 (\url{https://doi.org/10.54499/UIDP/00297/2020}) (Center for Mathematics and Applications) and the FCT scholarship, reference 2022.10596.BD (\url{https://doi.org/10.54499/2022.10596.BD}).
\bibliographystyle{abbrv}
\bibliography{biblio}
\section*{Appendix}
\subsection{Positive Turing machines and accepting states}\label{sec:no_accepting_states}
Recall that, in \cite{lautemann:96,lautemann:98}, positive Turing machines were defined having an accepting and rejecting state, whereas in \cite{das:18} only a \textit{halting} state is considered. We give an example as to why this change is required.

Suppose posDTM's had accepting and rejecting states. As \cite{lautemann:96,lautemann:98}, extend the order $\preceq_\frak{I}$ so that, if $I,I'\in\frak{I}$ and the state of $I'$ is accepting, then $I\preceq_\frak{I} I'$ (and analogously for the order $\preceq_\frak{O}$).
Then, one could define the following posDTM $M$:
\begin{center}
\begin{tikzpicture}[node distance = 3cm, on grid, auto]
    \node (init) at (-1,0) {\,};
    \node[state] (s) at (0,0) {$s$};
    \node[state] (q) at (3,0) {$q$};
    \node[state,accepting] (q_a) at (6,0) {$q_a$};
    \path[->]
    (init) edge (s)
    (s) edge [bend left] node {$\triangleright \to \triangleright,R$} (q)
    (q) edge [bend left] node {$0\to \square,L$} (s)
    (q) edge [bend left] node {$1\to \square,L$} (q_a)
    (q) edge [bend right] node[swap] {$\square \to 1,L$} (q_a);
\end{tikzpicture}  
\end{center}
Note that $M(0)=1$ and $M(1)=\eps$, so the size of the output does not depend only on the size of the input. 
Moreover, $M$ converts a distinction of $0$ and $1$ into a distinction on the \textit{size} of a word, which can then be computed by positive functions.
\subsection{The order on oracle Turing machine}\label{sec:oracle_order}
In Section \ref{sec:relativized_TM} we introduced positive oracle Turing machines, an extended the order on configurations by making the oracle states $q_{no}$ and $q_{yes}$ comparable. We give an example as to why this choice is necessary.

Suppose that we would leave $q_{no}$ and $q_{yes}$ incomparable, as the remaining states. Then, we could define the following positive oracle Turing machine:

\begin{center}
\begin{tikzpicture}[node distance = 3cm, on grid, auto]
    \node (init) at (-1,0) {\,};
    \node[state] (s) at (0,0) {$s$};
    \node[state] (q_q) at (3,0) {$q_{query}$};
    \node[state] (q_y) at (6,1) {$q_{yes}$};
    \node[state] (q_n) at (6,-1) {$q_{no}$};
    \node[state] (h) at (9,0) {$h$};
    \path[->]
    (init) edge (s)
    (s) edge node {$\triangleright \to \triangleright,R$} (q_q)
    (q_q) edge node[rotate=20,above,midway] {$x(i)\in A$} (q_y)
    (q_q) edge node[rotate=-20,below,midway] {$x(i)\notin A$} (q_n)
    (q_y) edge node[rotate=-20,above,midway] {$1\to 0,S$} (h)
    (q_n) edge node[rotate=20,below,midway] {$0\to 1,S$} (h);
\end{tikzpicture}  
\end{center}

Suppose $A=\{1\}$ and let the input be $x=1$.
After moving right from the $\triangleright$ symbol, the machine will enter the query state. Since $x\in A$, it will then enter the state $q_{yes}$, where it will convert the $1$ into a $0$. Analogously, if $x=0$, the machine will write a $1$ and halt.
Now, if $q_{no}$ and $q_{yes}$ are \textit{incomparable}, the above machine does not break the monotonicity restriction of the transition function, so, even relativized to a monotone oracle, the machine can compute a non-monotone function.

With the ordering on oracle Turing machines as defined in Definition \ref{def:positive_oracle_turing_machine}, this example is prevented.
If $q_{no}$ converts a $0$ into a $1$, it must also convert a $1$ into a $1$. Thus, as $q_{no}$ is smaller than $q_{yes}$, $q_{yes}$ must also convert a $1$ into a $1$.
\end{document}